\documentclass[11pt]{article}
\usepackage[mathscr]{eucal}
\usepackage{epsfig,amsfonts}
\usepackage{amsmath}
\usepackage{amsthm,amssymb}
\usepackage{graphicx}
\usepackage{hhline}
\usepackage{cite}
\usepackage{psfrag}
\usepackage{mathrsfs} 
\usepackage{hyperref}
\usepackage{bm}
\usepackage{array}
\usepackage{tikz}
\usepackage{multirow}
\usepackage{textcomp}
\DeclareMathAlphabet\mathbfcal{OMS}{cmsy}{b}{n}
\usepackage{hyperref}
\usepackage{mathrsfs} 
\usepackage{enumitem}
\usetikzlibrary{tikzmark,calc}
\DeclareMathAlphabet{\mathpzc}{OT1}{pzc}{m}{it}

\makeatletter
\@addtoreset{equation}{section}
\makeatother

\def\bea{\begin{eqnarray}}
	\def\eea{\end{eqnarray}}
\def\be{\begin{equation}}
	\def\ee{\end{equation}}

\def\be{\begin{equation}}
	\def\ee{\end{equation}}
\def\bdm{\begin{displaymath}}
	\def\edm{\end{displaymath}}
\def\bea{\begin{eqnarray}}
	\def\eea{\end{eqnarray}}

\def\ri{{\rm i}}

\def\XXint#1#2#3{{\setbox0=\hbox{$#1{#2#3}{\int}$}
		\vcenter{\hbox{$#2#3$}}\kern-.5\wd0}}

\newcommand{\p}{\partial}

\newcommand{\rd}{\mbox{d}}

\DeclareMathAlphabet{\mathpzc}{OT1}{pzc}{m}{it}

\begin{document}

\begin{titlepage}
$\phantom{I}$
\vspace{2.8cm}

\begin{center}
\begin{LARGE}

{\bf  Integrability and renormalizability for the fully anisotropic ${\rm SU}(2)$ principal
chiral field and its deformations}

\end{LARGE}

\vspace{1.3cm}
\begin{large}

{\bf 

Gleb A. Kotousov$^{1}$ and Daria A. Shabetnik$^{2}$}

\end{large}

\vspace{1.cm}
${}^{1}$Institut f$\ddot{{\rm u}}$r Theoretische Physik, 
Leibniz Universit$\ddot{{\rm a}}$t Hannover\\
Appelstra\ss e 2, 30167 Hannover, Germany\\
\vspace{.4cm}
${}^{2}$NHETC, Department of Physics and Astronomy\\
     Rutgers University\\
     Piscataway, NJ 08855-0849, USA\\
\vspace{1.0cm}

\end{center}

\begin{center}

\parbox{13cm}{%
\centerline{\bf Abstract} \vspace{.8cm}
For the class of $1+1$ dimensional field theories referred to as the
non-linear sigma models, there is known to be a  deep connection between classical integrability and
one-loop renormalizability. In this work, the phenomenon is reviewed
on the example of the so-called fully anisotropic ${\rm SU}(2)$ Principal Chiral Field (PCF).
Along the way, we discover
a new classically integrable four parameter family of sigma models, which is obtained from the
fully anisotropic ${\rm SU}(2)$ PCF by means of the
 Poisson-Lie deformation.
The theory turns out to be one-loop renormalizable
and the system of ODEs describing the flow of the four couplings
 is derived. Also provided are explicit analytical expressions for the full set of functionally 
independent first integrals (renormalization group invariants).
}
\end{center}

\vfill

\end{titlepage}

\setcounter{page}{2}

\tableofcontents

	\section{Introduction \label{sec1}}
	One of the spectacular instances of when ideas from physics and geometry come together is in the study of a class of field theories known as the Non Linear Sigma Models (NLSM). Mathematically, these are defined in terms of maps between two (pseudo-)Riemannian manifolds known as the worldsheet and the target space such that the classical equations of motion take the form of a generalized version of Laplace's equation
	\cite{Fuller}.   In physics, one of the uses of NLSM is as low energy effective field theories with the choice of the target space being dictated by the symmetries of the problem. The first such proposal appeared in a paper of  Gell-Mann and Levy \cite{G-L}. They put forward the following Lagrangian density as an  effective field theory of pions:
	\be
	{\cal L}=\frac{1}{2}\, \eta^{ij}\,\p_i \vec n\cdot\p_j \vec n \,\qquad {\rm with}
	\qquad |\vec n|^2=\frac{1}{f^2}\,.
	\label{LGL}
	\ee
	Here the last equation means that the four component field $\vec n=(n_1, n_2, n_3, n_4)$ is constrained  to lie on the three dimensional round sphere  whose radius coincides with $1/f$. Thus the target space is ${\mathbb S}^3$ equipped with the homogeneous metric  while the worldsheet is four dimensional Minkowski spacetime $\mathbb{M}^{1,3}$.  The field theory is known as the  ${\rm O}(4)$ sigma model as it possesses ${\rm O}(4)$ symmetry -- the group of isometries of the three-sphere. Ignoring global aspects, one may replace the latter   by  ${\rm SU}(2)\times {\rm SU}(2)$  which  play the r{o}le of the vector and axial symmetries appearing in the `chiral limit' of QCD.
	For this reason the model \eqref{LGL} is also referred to as the ${\rm SU}(2)$ principal chiral field.
	\bigskip
	
	The ${\rm O}(4)$ sigma model is rather special in
	$1+1$  dimensional spacetime $\mathbb{M}^{1,1}$. In this case, as was pointed out by Polyakov,  the Lagrangian \eqref{LGL} defines a renormalizable QFT. Following the traditional path-integral quantization, the model should be  equipped with   a UV cutoff $\Lambda$
\cite{Polyakov}. 
It was shown to one-loop order that  a consistent removal of the UV divergences can be achieved if the bare coupling is given a dependence  on the cutoff momentum, described by the RG flow equation  \cite{Polyakov1}
	\be
	\Lambda\frac{{\rd}}{{\rd}\Lambda} (f^{-2})=\frac{N-2}{2\pi}\,\hbar+O(\hbar^2)\,.
	\label{R}
	\ee
	Here $\hbar$ stands for the   dimensionless Planck constant while $N=4$ (the computation was performed for the  general ${\rm O}(N)$ sigma model with  target space $\mathbb S^{N-1}$). Notice that in the continuous limit $\Lambda\to \infty$ the coupling constant $f^2$ approaches zero. In turn, the curvature of the sphere to which the fields $n_j(x^0,x^1)$ belong vanishes so that the theory becomes non-interacting.  This phenomenon, known as asymptotic freedom, indicates consistency of the quantum field theory. As a result of  the work of Polyakov and later Zamolodchikov and Zamolodchikov \cite{ZZ}, who proposed the associated scattering theory, it is commonly believed  that  the ${\rm O}(N)$ sigma model in $1+1$  dimensions  is a well defined  
	(UV complete) QFT.
	\bigskip

	The renormalizability of general NLSM  in $1+1$ dimensions was discussed in the work of Friedan \cite{Friedan}. He considered the class of theories where the Lagrangian density takes the form
	\be
	\label{Ls}
	\mathcal{L}=\frac{1}{2}\,G_{\mu\nu}(X)\,\eta^{ij}\p_i X^\mu\p_j X^\nu\,.
	\ee
	Here $G_{\mu\nu}(X)$ is the metric written  in terms of  local coordinates $X^\mu$  on the target space.
	The couplings are encoded in  this metric so that the latter is taken to be dependent on the cutoff $\Lambda$.
	Extending the results of Ecker and Honerkamp \cite{EH}, Friedan computed the RG  flow equation  to two loops. To the leading order in $\hbar$ it takes the form 
	\be\label{ricci}
	\p_\tau G_{\mu\nu}=-\hbar\,R_{\mu\nu}+O(\hbar^2)\,,\qquad\qquad \p_\tau=-2\pi \Lambda\,\frac{\p}{\p\Lambda}\,,
	\ee
	where $R_{\mu\nu}$ is the Ricci tensor built from the metric.  Without the  $O(\hbar^2)$ term, \eqref{ricci}  is  usually referred to  as  the Ricci flow equation \cite{H}, which is a partial differential equation for 
	$G_{\mu\nu}=G_{\mu\nu}(X\,|\,\tau)$. It found a remarkable application in mathematics in the proof of the 
	Poincar\'{e} conjecture \cite{Perelman1,Perelman2}.
	\bigskip

	The question of renormalizability can be addressed within a class of NLSM where the target space metric
	depends on a  finite number of parameters. 
	The simplest example is the ${\rm O}(N)$ sigma model whose target manifold belongs to the family
	of  the $(N-1)$ dimensional round spheres,  characterized by the radius $1/f$. 
	In this case, the Ricci flow equation  boils down to the ordinary differential equation \eqref{R}.
	Another  example is the Principal Chiral Field (PCF), where the target space is the group manifold
	of a simple Lie group $G$ equipped with the  left/right invariant metric. The latter is
	unique up to homothety and, in local coordinates,  is defined by the relation
	\be 	\label{L2}
	G_{\mu\nu}(X)\,{\rd}X^\mu{\rd}X^\nu=-\frac{1}{e^2}\,
	\big\langle {\bm U}^{-1}\,{\rd {\bm U}}\,,{\bm U}^{-1}\,{\rd {\bm U}}\big\rangle\ ,
	\ee
	where  $\bm{U}\in G$,  $e$ is the homothety parameter and the angular brackets $\langle\cdot, \cdot\rangle$ denote the Killing form in the Lie algebra of $G$.\footnote{For a classical Lie group we take the Killing form 
		to be the trace over the defining representation.}
	The Ricci flow  \eqref{ricci} implies
	\be\label{san121}
	\p_\tau (e^{-2})=-\tfrac{1}{2}\,C_2\,\hbar+O(\hbar^2)\,
	\ee
	with $C_2$ being the value of the quadratic Casimir in the adjoint representation.  
	This equation was essentially obtained in the original work of Polyakov \cite{Polyakov1}, see also \cite{Polyakov}.
	Notice that the ${\rm SU}(2)$ PCF coincides with the ${\rm O}(4)$ sigma model. 
	In this case $C_2=2$, while \eqref{san121} and \eqref{R} are the same
	provided that $e^2\equiv 2f^2$.

	\bigskip

	An example of an NLSM which is  renormalizable within a two parametric family is the so-called anisotropic ${\rm SU}(2)$  PCF. In this case the ${\rm SU}(2)\times {\rm SU}(2)$ isometry of the target space is broken down to ${\rm SU}(2)\times{\rm U}(1)$ and  the 
	manifold is still topologically ${\mathbb S}^3$ but equipped with a certain asymmetric metric.  The latter is given by
	\be
	G_{\mu\nu}(X)\,{\rd}X^\mu{\rd}X^\nu=-\frac{1}{e^2}\,\big\langle {\bm U}^{-1}\,{\rd}{\bm U}\,,{\cal O}( {\bm U}^{-1}\,{\rd}{\bm U})\big\rangle\,,
	\label{apfc}
	\ee
	where ${\cal O}$ is an operator acting from the Lie algebra $\mathfrak{su}(2)$ to itself depending on the additional deformation parameter $r$,
	\be
	{\cal O}:\ {\mathfrak{su}(2)}\mapsto  {\mathfrak{su}(2)}\,,\qquad\qquad\qquad{\cal O}=1+r\,P_3\,,
	\ee
	and $P_3$ projects onto the Cartan subalgebra.
	The Ricci flow equation reduces to a system of ordinary differential equations on $e$ and $r$:
	\bea\label{sa9832hjd}
	-\frac{1}{\hbar}\,\p_\tau(e^{-2})&=&1-r\nonumber\\[0.2cm]
	-\frac{1}{\hbar}\,\p_\tau r &=&2\,e^2 r \,(r+1)\,.
	\eea
	In the domain $-1<r<0$, similar as with the ${\rm SU}(2)$ PCF,
	the theory is asymptotically free and it turns out to be a consistent QFT.
	
	\bigskip
	When the $\tau$ dependence of the metric, satisfying the Ricci flow equation, is contained
	in a finite number of parameters, the partial differential equation \eqref{ricci} reduces to a system of ordinary
	ones. From the point of view of physics, this means that the corresponding 
	NLSM depends on a finite number of coupling constants and 
	is one-loop renormalizable within this class. The construction of such solutions is difficult to achieve even
	when the dimension of the target manifold is low. Among the most impressive early results was the work 
	of Fateev \cite{Fateev}, who discovered a three parameter family of metrics solving the Ricci flow
	equation. The NLSM with this background is a two parameter deformation of the ${\rm SU}(2)$ PCF,
	which contains the anisotropic case as a subfamily. A guiding principle for exploring the class of renormalizable NLSM
	was formulated in the work \cite{Lukyanov}. It  arose from the observation that all
	the above mentioned   models  turn out to be classically integrable field theories.
	It is now believed that there is a deep relation between classical integrability and one-loop renormalizability in 
	$1+1$ dimensional sigma models.  
	\bigskip

\begin{figure}
		\begin{center}
			\begin{tikzpicture}
				\draw (0,0) ellipse (1.25 and 0.5);
				\draw (-1.25,0) -- (-1.25,-3.5);
				\draw (-1.25,-3.5) arc (180:360:1.25 and 0.5);
				\draw (-1.25,-2) arc (180:360:1.25 and 0.5);
				\draw (-1.25,-1.2) arc (180:360:1.25 and 0.5);
				\draw [dashed] (-1.25,-3.5) arc (180:360:1.25 and -0.5);
				\draw [dashed] (-1.25,-2) arc (180:360:1.25 and -0.5);
				\draw [dashed] (-1.25,-1.2) arc (180:360:1.25 and -0.5);
				\draw (1.25,-3.5) -- (1.25,0);  
				\fill [gray,opacity=0.2] (-1.25,0) -- (-1.25,-3.5) arc (180:360:1.25 and 0.5) -- (1.25,0) arc (0:180:1.25 and -0.5);
				\draw [->] (-2.25, -3) -- (-2.25, -1);
				\node[text width=0.5cm] at (-2.5, -2) {$x^0$};
				\node[text width=2cm] at (2.5
				,-2) {$x^1\sim x^1+R$};
			\end{tikzpicture}
			\caption{\small%
The integration contour for the Wilson loop can be moved freely along the cylinder.}
		\end{center}
		\label{marker1}
	\end{figure}
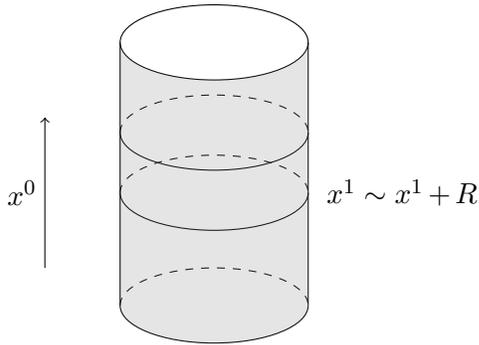

	The notion of classical integrability in $1+1$ dimensional field theory requires explanation. 
	Recall that a mechanical system  with $d$ degrees of freedom is called  integrable (in the Liouville sense)
	if it possesses $d$  functionally independent   Integrals of Motion (IM) in involution. This concept is difficult to extend to a field theory, where the  number of degrees of freedom is infinite. A suitable paradigm of integrability in the case of $1+1$ dimensions  arose  from the works of the Princeton group \cite{Pr} and was later developed in the papers of Lax \cite{Lax} and Zakharov and Shabat \cite{ZS}. 
	A key ingredient is the existence of the so-called Zero Curvature Representation  (ZCR) of the  Euler-Lagrange equations
	of the classical field theory:
	\be
	[\p_i-{\bm A}_i,\ \p_j-{\bm A}_j]=0\,.
	\label{ZCR}
	\ee
	Here ${\bm A}_{i}={\bm A}_{i}(x^0,x^1|\lambda)$ is a Lie-algebra valued worldsheet connection which also depends on the auxiliary (spectral) parameter $\lambda$. 
	The ZCR implies that the Wilson loops
	\be
	T(\lambda)={\rm Tr}\ \overset{\leftarrow}{\cal P}\exp\bigg(-\int_{\cal C} {\rd} x^i\,{\bm A}_i\bigg)\,,
	\label{T}
	\ee 
	where the trace is taken over some matrix representation of the Lie algebra, are unchanged under
	continuous deformations of the closed contour ${\cal C}$. If suitable  boundary conditions are imposed, this can be used to generate IM. For instance, in the case when the worldsheet is the cylinder and the connection
	is single valued,  the contour    ${\cal C}$ may be chosen to be the equal-time slice at some $x^0$ as in 
	fig.\,\ref{marker1}. Then, it is easy to see  that $T(\lambda)$ does not depend on the choice of $x^0$, i.e.,
	it is an integral of motion. Due to
	the dependence on the arbitrary complex variable $\lambda$, $T(\lambda)$
	constitutes a  family of IM. The existence of these may provide a 
	starting point for solving  the classical equations of motion by applying the inverse scattering transform \cite{Faddeev}. For this reason, we say that a $1+1$ dimensional classical field theory is integrable if it admits the ZCR.\footnote{Such a `definition' of integrability does not guarantee that the equations of motion can be analytically solved in any sense.
		Thus, it is a much weaker notion than Liouville integrability in classical mechanics.}  
	\bigskip
	
	The theme of this paper is the interplay between  classical integrability and one-loop renormalizability in sigma models.
           Its  structure is as follows.
	Section \ref{sec2} is devoted to a discussion of the so-called fully anisotropic ${\rm SU}(2)$ PCF, whose target space metric is given by
	\be
	G_{\mu\nu}(X)\,{\rd}X^\mu{\rd}X^\nu=- 2\big\langle {\bm U}^{-1}\,{\rd}{\bm U}\,,{\cal O}( {\bm U}^{-1}\,{\rd}{\bm U})\big\rangle\,,\qquad\qquad {\cal O}=I_1\,P_1+I_2\,P_2+I_3\,P_3\,.
	\label{mm1}
	\ee
	Here $P_a$ are  projectors onto the basis ${\tt t}_a$ of the Lie algebra $\mathfrak{su}(2)$, which is taken to be orthogonal w.r.t. the Killing form. 
	The theory is a two parameter deformation of the ${\rm SU}(2)$ PCF and it reduces to the latter when $I_1=I_2=I_3=\frac{1}{2}\,e^{-2}$. In addition for the special case
	$I_1=I_2$ it becomes the anisotropic ${\rm SU}(2)$ PCF, whose  target space metric was presented above in
	eq.\,\eqref{apfc}. We discuss the classical integrability of the model with metric \eqref{mm1}.
	On the other hand, the latter is shown  to be a
	solution of the Ricci flow equation for a certain $\tau$ dependence of the
	couplings $I_a=I_a(\tau)$. The corresponding system of ordinary differential equations is derived 
	and its first integrals are obtained.
	In section \ref{sec3} the concept of the Poisson-Lie deformation \cite{Klimchick3}, which preserves integrability, is introduced.
          We apply it to the  fully anisotropic ${\rm SU}(2)$ PCF and obtain a new classically integrable field theory depending on four parameters.
	It is argued that the resulting model is  one-loop renormalizable. 
The system of ODEs
	for the $\tau$ dependence of the four couplings is presented and 
explicit analytical expressions for the
renormalization group invariants are provided. The last section is devoted to a discussion. Among other things, it contains the formulae for the
renormalization group invariants of the fully anisotropic ${\rm SU}(2)$ PCF with Wess-Zumino term.

	\section{Fully anisotropic ${\rm SU}(2)$ PCF\label{sec2}}
Following the lecture notes \cite{Lukyanov:2019asr}, let us gain some intuition about the fully anisotropic ${\rm SU}(2)$ PCF by 
considering its classical mechanics counterpart. 
It is obtained via   `dimensional reduction' where one restricts to  
field configurations that depend only on the spacetime variable $x^0$ 
so that ${\bm U}={\bm U}(x^0)$. Then the Lagrangian  density \eqref{Ls}, \eqref{mm1} becomes 
	\be
	L=\sum\limits_{a=1}^3\frac{I_a\,\omega_a^2}{2}\,,
	\label{l4}
	\ee
	where $\omega_a$ are defined through the relation
	\be
	{\bm U}^{-1}\,\dot {\bm  U}=-\ri\,\sum\limits_{a=1}^3\omega_a\, {\tt t}_a\,
	\label{jj}
	\ee
	and the dot stands for  differentiation w.r.t. the time $x^0$. Also, the basis for the Lie algebra has been normalized such that 
	\be\label{sakj78hjsghaaad21}
	\langle {\tt t}_a, {\tt t}_b \rangle=\tfrac{1}{2}\,\delta_{ab} \qquad\qquad {\rm and}\qquad\qquad
	[ {\tt t}_a, {\tt t}_b]=\ri\, \epsilon_{abc} {\tt t}_c 
	\ee
	with  $\epsilon_{abc}$ being the Levi-Civita symbol and summation over the repeated index is being assumed. It turns out that 
	the Lagrangian \eqref{l4} describes the free motion of a rigid body where
	the  translational degrees of freedom   have been  ignored. 
	\bigskip
	
	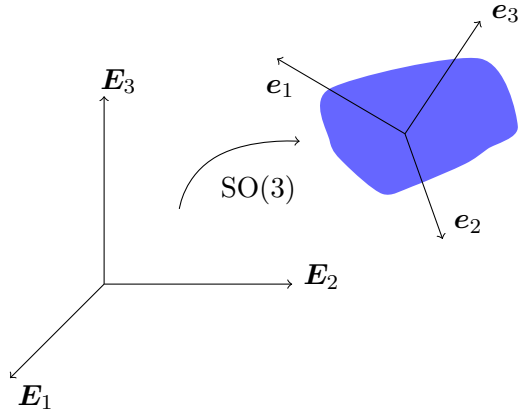
\begin{figure}
		\begin{center}			
\begin{tikzpicture}
				\draw [->] (0,0)--(0,2.5);
				\draw [->] (0,0)--(2.5,0);
				\draw [->] (0,0)--(-1.25, -1.25);
				\node[text width=0.5 cm] at (0.2,2.7) {${\bm E}_3$};
				\node[text width=0.5 cm] at (-0.9,-1.5) {${\bm E}_1$};
				\node[text width=0.5 cm] at (2.9,0.1) {${\bm E}_2$};
				\node[text width=0.5 cm] at (5.4, 3.6) {${\bm e}_3$};
				\node[text width=0.5 cm] at (4.9, 0.8) {${\bm e}_2$};
				\node[text width=0.5 cm] at (2.4, 2.6) {${\bm e}_1$};
				\fill [blue!60!]  plot [smooth cycle,tension=0.5] coordinates {(5,3) (5.5, 2.1)(5.1,1.8) (4.8,1.6) (4.1, 1.3)(3.7,1.2) (3.2,1.6)(3.0,1.9)(3.0,2.6)};
				\draw [->]  (4.0, 2)--(5, 3.5);
				\draw [->]  (4.0, 2)--(4.5, 0.6);
				\draw [->]  (4.0, 2)--(2.3,3);
				\draw [->] plot [smooth, tension=1] coordinates { (1,1)  (1.5,1.7)(2.6,1.9) };
				\node[text width=0.5 cm] at (1.8, 1.25) {${\rm SO}(3)$};
			\end{tikzpicture}
			\caption{\small%
The orientation of the rigid body  is uniquely specified by the $3D$ special orthogonal matrix that relates    the moving  frame  
				$({\bm e}_1,\bm{e}_2,\bm{e}_3)$ to the fixed frame $({\bm E}_1,\bm{E}_2,\bm{E}_3)$. The axes of the moving frame are chosen to coincide with the principal axes of inertia.}
		\end{center}
		\label{marker}
	\end{figure}
	
	Recall that an arbitrary displacement of a rigid body is a composition of a translation and a  rotation.  
	For a free moving top, when the net external force is zero, one can without loss of generality consider the case when the centre of mass is at rest.
	Introduce two right handed coordinate systems called the fixed (laboratory) frame and moving frame, which are defined by the ordered set of unit vectors
	$(\bm{E}_1,\bm{E}_2,\bm{E}_3)$ and $(\bm{e}_1,\bm{e}_2,\bm{e}_3)$, respectively. The axes of the moving frame coincide with
	the principal axes of the rigid body w.r.t. the centre of mass.
	Then  the orientation of the body is uniquely specified by a $3\times 3$ special orthogonal matrix which relates the  fixed and moving frames  as in fig.\,\ref{marker}. 
	Thus the configuration space of a rigid body with a fixed point  coincides with the group manifold of $ {\rm SO}(3)$.
	The  matrix specifying the rotation  can be identified with an ${\rm SU}(2)$ matrix ${\bm U}$ taken in the adjoint representation. Mathematically this  is expressed as
	\be
	{\bm E}_a{\tt t}_a={\bm U}\, {\bm e}_a{\tt t}_a\,{\bm U}^{-1}\,,
	\label{wde}
	\ee
	where again summation over $a=1,2,3$ is being assumed.
	The coefficients $\omega_a$ defined in \eqref{jj} coincide with the projections of the instantaneous angular velocity ${\bm \omega}$ along the principal axes. This can be seen by differentiating 
	both sides of \eqref{wde}  w.r.t. time and comparing the result with $\dot{\bm e}_a={\bm \omega}\times{\bm e}_a$. 
	\bigskip
	
	The classical mechanics system governed by
	the Lagrangian  \eqref{l4} is called the Euler top. The parameters $I_a$, which were introduced originally  as formal couplings in  \eqref{mm1}, coincide with the  principal moments of inertia. 
	Notice that the Lagrangian is built from $\bm{U}^{-1}\dot{\bm{U}}$ which belongs to the Lie algebra and hence is insensitive to the difference between the groups
	${\rm SU}(2)$ and ${\rm SO}(3)$.\footnote{%
		Topologically, the special unitary group ${\rm SU}(2)$ is the three sphere $\mathbb{S}^3$, 
		while the special orthogonal group ${\rm SO}(3)$ is the three dimensional real projective space $\mathbb{RP}^3$.
		The latter coincides with $\mathbb{S}^3$ with antipodal points $\pm\vec{n}\in\mathbb{S}^3$ identified.}
	
	\bigskip

	The Euler top   is  a textbook example of a Liouville integrable system. The  IM that satisfy the conditions of Liouville's theorem are the Hamiltonian $H$  and two more which are built from the angular momentum ${\bm M}$:
	\be
	H=\sum\limits_{a=1}^3 \frac{I_a \, \omega_a^2}{2}\,,\qquad\qquad {\bm M}=\sum_{a=1}^3I_a\omega_a\, {\bm e}_a\,.
	\ee
	For a free moving body the angular momentum is conserved, i.e., $\dot{\bm{M}}=0$. On the other hand, the total time derivative 
	$\dot{\bm{M}}$  can be written in terms of the canonical Poisson bracket as 
	$\{H,\bm{M}\}$. Hence, the classical observable $\bm{M}$ Poisson commutes with the Hamiltonian. This way, the three functionally independent involutive Integrals of Motion may be taken to be
	\be
	H\,, \qquad\qquad M_Z\equiv{\bm M}\cdot {\bm E}_3\,\qquad\qquad {\rm and}\qquad\qquad {\bm M}^2=\sum_{a=1}^3I_a^2\, \omega_a^2\,.
	\ee
	It  follows from   Liouville's theorem that the equations of motion for the Euler top can be integrated in quadratures. The
	solution is discussed in any standard textbook on classical mechanics  see, e.g.,  \cite{Landau}.
	\bigskip
	
	The rigid body with  two of the principal moments of inertia equal $I_1=I_2\equiv I$ is 
	usually referred to as the symmetric top. In this case  the Lagrangian \eqref{l4} possesses invariance w.r.t. rotations about the  axis ${\bm e}_3$. 
	For the symmetric top  it is convenient to choose the three functionally independent, involutive  IM to be ${\bm M}^2, M_Z$ and $M_3\equiv{\bm M}\cdot {\bm e}_3$. 
	Notice that the Hamiltonian is given in terms of these as 
	\be
	H=\frac{1}{2I}\,\bm{M}^2+\Big(\frac{1}{2I_3}-\frac{1}{2I}\Big)\,M_3^2\ \qquad\qquad\qquad (I_1=I_2\equiv I)\,.
	\ee
	The case $I_1=I_2=I_3\equiv I$  
	is known as the spherical top and the Hamiltonian is proportional to $\bm{M}^2$.
	The field theory generalization of the symmetric top is the anisotropic ${\rm SU}(2)$ PCF \eqref{Ls},\,\eqref{apfc},
	while that of the spherical top is the ${\rm SU}(2)$ PCF \eqref{Ls},\,\eqref{L2}.

	\bigskip
	
	Remarkably,   the fully anisotropic ${\rm SU}(2)$ PCF is also an  integrable field theory  according to the technical  definition given in the introduction. Namely, the equations of motion for the model admit the 
	Zero Curvature Representation \eqref{ZCR}. To demonstrate the integrability, it is useful to introduce the currents $J_i^a$ via the formula:
	\be
	{\bm U}^{-1}\,\p_{i} {\bm U}=-\ri\,\sum\limits_{a=1}^3 J_{i}^a\, {\tt t}_a\,\qquad\qquad\qquad \qquad (i=0,1)\,.
	\label{re}
	\ee
	Then the Euler-Lagrange equations for the model \eqref{Ls}, \eqref{mm1} can be written as follows:
	\be	\label{eqr}
	\p_-J_+^a+\p_+J_-^a=\frac{I_b-I_c}{I_a}\ (J_+^b\,J_-^c+J_+^c\,J_-^b)\,,
	\ee
	where $(a,b,c)$ is a cyclic permutation of $(1,2,3)$ while 
	\be
	\p_{\pm}=\tfrac{1}{2}\,(\p_0\pm\p_1)\,,\qquad\qquad\qquad J_\pm^a=\tfrac{1}{2}\,(J_0^a\pm J_1^a)\,.
	\ee
	Note also the kinematic relations (Bianchi identities) which follow directly from the definition \eqref{re}:
	\be	\label{aaa}
	\p_-J_+^a-\p_+J_-^a=\epsilon_{abc}\,J_+^bJ_-^c\,.
	\ee 
	The worldsheet connection  for the fully anisotropic ${\rm SU}(2)$ PCF is rather complicated.
	For this reason we give it first for the case  $I_1=I_2=I_3=\frac{1}{2}\,e^{-2}$ which corresponds to the ${\rm SU}(2)$ PCF. Then the
	equations of motion \eqref{eqr} simplify greatly since the term in the r.h.s. vanishes. The worldsheet connection ${\bm A}_{\pm}$ reads as
	\be
	{\bm A}_{\pm}=\frac{\ri\,J_\pm^a\,{\tt t}_a}{1\pm \lambda}\qquad\qquad (I_1=I_2=I_3)\,
	\label{zcr11}
	\ee
	and one can easily check that as a consequence of eqs.\,\eqref{eqr} and \eqref{aaa},
	\be
	[\p_+-{\bm A}_+,\ \p_--{\bm A}_-]=0\,.
	\ee
	This ZCR was first proposed  in the work \cite{ZM} and is valid for the sigma model associated with any simple Lie group $G$ with $\ri J_\pm^a\,{\tt t}_a$ replaced by $-{\bm U}^{-1}\,\p_\pm {\bm U}$.
	\bigskip
	
	The ZCR for the general case with $I_1\neq I_2\neq I_3$   was found in \cite{Cerednik} and  presented in a
slightly different form in ref.\cite{EllipticLax}. 
	In the following, the conventions of the latter paper will be used.
	To write the result, we swap the two independent combinations of $(I_1, I_2, I_3)$ that enter into the equations of motion for $m$ and $\nu$ according to
	\be
	m=\frac{I_2\,(I_1-I_3)}{I_3\,(I_1-I_2)}\,,\qquad\qquad {\text{cn}}^2(\nu,m)=\frac{I_1}{I_2}\,,
	\label{mpm}
	\ee
	where $\text{cn}(\nu,m)$ is the Jacobi elliptic function with \emph{the parameter}  $m$. Together with ${\rm sn}$ and ${\rm dn}$, it satisfies the relations
	\be
	{\rm sn}^2(\nu,m)+{\rm cn}^2(\nu,m)=1\,,\qquad\qquad m\,{\rm sn}^2(\nu,m)+{\rm dn}^2(\nu,m)=1\,.
	\ee
	The flat worldsheet connection  reads explicitly as
	\be
	{\bm A}_\pm=\ri\,\sum\limits_{a=1}^3w_a(\nu\mp\lambda)J_\pm ^a\, {\tt t}_a\,,
	\label{wwwew}
	\ee
	where 
	\be
	w_1(\lambda)= \frac{{\rm sn}(\nu,m)}{\text{sn}(\lambda,m)}\,,\qquad 
	w_2(\lambda)=\frac{{\rm sn}(\nu,m)}{{\rm cn}(\nu,m)} \frac{\text{cn}(\lambda,m)}{\text{sn}(\lambda,m)}\,,\qquad   \  w_3(\lambda)=\frac{{\rm sn}(\nu,m)}{{\rm dn}(\nu,m)} \frac{\text{dn}(\lambda,m)}{\text{sn}(\lambda,m)}\,.
	\label{wwwww}
	\ee

	\bigskip
	
	In order to explore the one-loop renormalizability of the fully anisotropic ${\rm SU}(2)$ PCF, we turn to the analysis of the Ricci flow equation \eqref{ricci}. It requires one to calculate   the Ricci tensor $R_{\mu\nu}$  corresponding to the target space metric  $G_{\mu\nu}$ given in \eqref{mm1}. The computation is straightforward and we do not present it here. Instead, we mention the identity: 
	\be\label{akjs87hjd}
	R_{\mu\nu}=\sum\limits_{a=1}^3\frac{(I_a-I_b+I_c)(I_a+I_b-I_c)}{2I_bI_c}\,\frac{\p}{\p I_a}\, G_{\mu\nu}\,,
	\ee
	where $(a,b,c)$ is a cyclic permutation of $(1,2,3)$. Then it follows that the Ricci flow equation is satisfied   if the couplings  $I_a$ are assigned a $\tau$ dependence such that (see also refs.\cite{Bakas:2006bz,Sfetsos:2014jfa})
	\be\label{sakj39821}
	-\frac{1}{\hbar}\,\p_{\tau}(I_a\,I_b)=I_a+I_b-I_c\,,\qquad\qquad (a,b,c)={\tt perm}(1,2,3)\, .
	\ee
	This constitutes a  set of coupled nonlinear ordinary differential equations describing the flow. 
           Notice that for $I_1=I_2=\frac{1}{2e^2}$ and 
           $I_3=\frac{1+r}{2e^2}$ one recovers the Ricci flow equations for the anisotropic ${\rm SU}(2)$ PCF \eqref{sa9832hjd}.
          The latter reduce to the ones for the ${\rm SU}(2)$ PCF \eqref{san121} with $C_2=2$ upon setting $r=0$.
           
	\bigskip
	
	We found that the system \eqref{sakj39821} possesses two Liouvillian first integrals.\footnote{%
		Liouvillian first integrals are those that are expressed in quadratures in the dependent variables of the differential equation.}
	They are given by
	\be\label{an8973jsd}
	Q_1=\frac{K(1-m)-(1-p) E(1-m)}{(1-p) E(m)+p K(m)}\,,\qquad\qquad   
Q_2=\frac{I^2_1\,\big((p-1)\, E(m)-p\, K(m) \big)^2}{p\,(p-1)\,  (p\, m-m+1)}\,.
	\ee
	Here $K(m)$ and $E(m)$ stand for the complete elliptic integrals of the first and second kind,
	\be
	K(m)=\int^{\frac{\pi}{2}}_0 \frac{{\rd }\theta}{\sqrt{1-m\, \sin^2\theta}}\,,\qquad\qquad E(m)=\int^{\frac{\pi}{2}}_0\,{\rd }\theta\, \sqrt{1-m\, \sin^2\theta}\,,
	\ee
	the parameter $m$ is the same as in \eqref{mpm}, while $p$ coincides with ${\text{cn}}^2(\nu,m)$ from that formula, i.e.,
	\be\label{an8973jsdC}
	m=\frac{I_2\,(I_1-I_3)}{I_3\,(I_1-I_2)}\,,\qquad\qquad\qquad
	p=\frac{I_1}{I_2}\,.
	\ee
	The expression \eqref{an8973jsd} for the first integrals is one of the original results of this paper.\footnote{%
A set of first integrals of the system \eqref{sakj39821}, similar to  \eqref{an8973jsd},
have also appeared in the recent work \cite{Lacroix:2024wrd}. Their results and the ones of our study 
were achieved independently of each other.}   
After it was obtained, we discovered
	that the system of differential equations \eqref{sakj39821}
	had  been introduced, in a slightly different form, in the work of Darboux\,\cite{Darboux}. Its solution
	was discussed in  refs.\cite{Halphen1,Halphen2}. 
          
	\bigskip

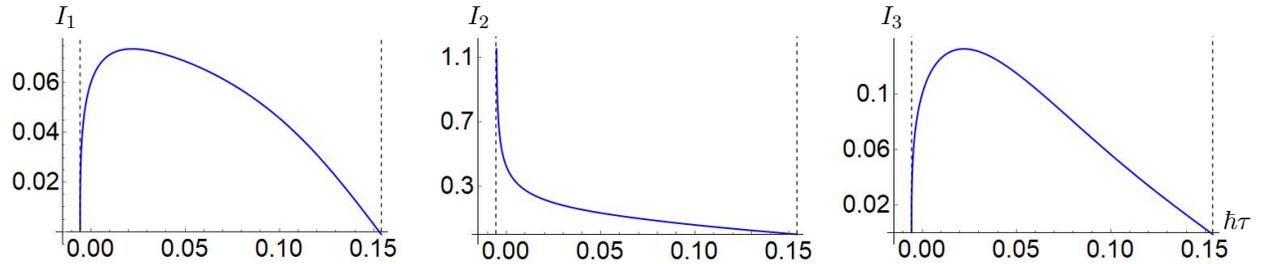
\begin{figure}[h!]
			\centering
			\scalebox{0.9}{
				\begin{tikzpicture}
					\node at (-1.5,0)
					{\includegraphics[width=1\linewidth]{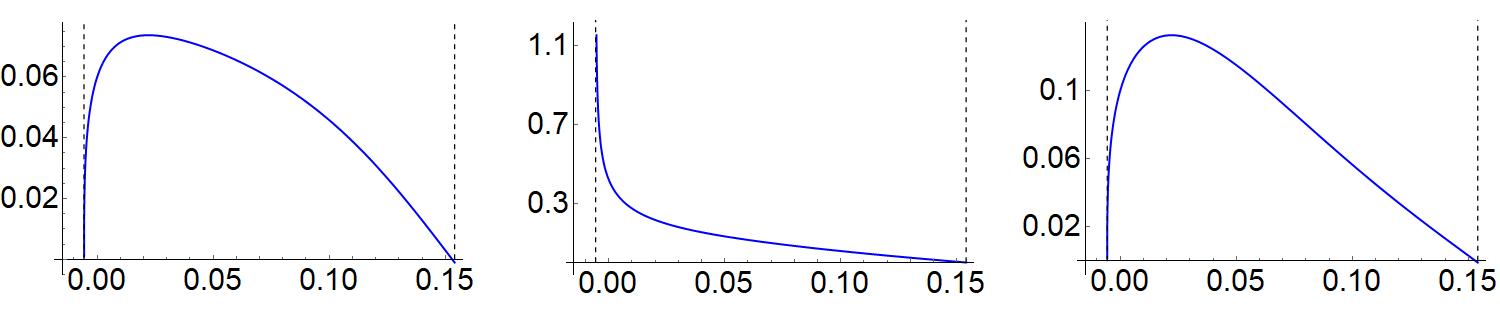}};
					\node[] at  (7.6,-1) {$\hbar\tau$};
					\node[] at  (2.5,2) {$I_3$};
					\node[] at  (-3.6,2) {$I_2$};
					\node[] at  (-9.7,2) {$I_1$};
			\end{tikzpicture}}
			\caption{\small%
The evolution of $I_1,\, I_2$ and $I_3$ as functions of $\hbar\tau$. The initial conditions at $\tau=0$ were chosen to be $I_1(0)=0.06,\ I_2(0)=0.42,\ I_3(0)=0.10$. The flow remains real and non-singular in the interval
$\tau\in (\tau_{\text{min}}, \tau_{\text{max}})$ with $  \hbar\tau_{\text{min}}=-0.006$ 
and $\hbar\tau_{\text{max}}=0.154$ which is marked by the dashed lines.
				}
			\label{f3}
		\end{figure}

	The flow of the couplings $I_a$ as a function of $\tau$  can be analyzed numerically. The typical behaviour, for generic initial conditions such that all $I_a$ at $\tau=0$ are positive and different, 
	is presented in  fig.\,\ref{f3}. One observes from the figure that 
	the solution of  \eqref{sakj39821}, i.e., the Ricci flow equation,
	remains real and non-singular only within the finite interval $\tau\in(\tau_{\rm min},\tau_{\rm max})$. At the end points  one of the couplings goes to zero so that the curvature of the target space    blows up. As a result,  the one-loop approximation is no longer valid and the perturbative analysis is not sufficient
	to explore whether or not the model  can be defined as a consistent (UV complete) QFT. 

		\bigskip

		There exists another three parameter family of deformations of the  three dimensional round sphere \eqref{L2}.
		It is the one mentioned in the introduction that was proposed by Fateev in ref. \cite{Fateev}. His metric, depending on  $(e^2,r,l)$, can be written  as
		\be\label{i389dw87}
		G_{\mu\nu}(X)\,{\rd}X^\mu{\rd}X^\nu=-\frac{(1+r)(1+l)}{e^2}\ \frac{
			\big\langle {\bm U}^{-1}\,{\rd}{\bm U}\, ,\, {\cal O}({\bm U}^{-1}\,{\rd}{\bm U})\big\rangle}{(1+r)(1+l)-4rl\, \big(\langle {\bm U}{\tt t}_3 {\bm U}^{-1}, {\tt t}_3\rangle\big)^2}\,.
		\ee
		Here the operator ${\cal O}$, acting on the Lie algebra, is given by
		\be\label{i389dw87AA}
		{\cal O}=1+r\,P_3+l\,{\rm Ad}_{\bm U}\circ P_3\circ {\rm Ad}^{-1}_{{\bm U}}\,,
		\ee
		where ${\rm Ad}_{\bm U}$ stands for the adjoint action of the group: 
		\be
		{\rm Ad}_{\bm U}\, {\tt x}={\bm U}^{-1}{\tt x}\,{\bm U}\,, \qquad\qquad {\tt x}\in {\mathfrak{su}}(2)\,.
		\label{adj}
		\ee
		The Ricci flow equation  \eqref{ricci} leads to the system of ordinary differential equations for the three parameters:
  \bea\label{rff}
-\frac{1}{\hbar}\,\p_\tau l&=&\frac{2\,e^2\, l \,  (1+l+r)}{1+r}\nonumber\\[0.2cm]-\frac{1}{\hbar}\,\p_\tau r&=&\frac{2\,e^2\, r \,  (1+l+r)}{1+l}\\[0.2cm]
-\frac{1}{\hbar}\,\p_\tau(e^{-2})&=&\frac{(1-l-r) (1+l+r)}{(1+l) (1+r)}\ .\nonumber
  \eea
		Notice that for $l=0$ the metric  \eqref{i389dw87}  becomes the one for the anisotropic ${\rm SU}(2)$ PCF \eqref{apfc}, 
		while the above system of differential equations reduces to \eqref{sa9832hjd}.
\smallskip

		A remarkable feature of \eqref{rff}  is that it possesses solutions where $e^{2}(\tau),\, r(\tau),\, l(\tau)$
		are real and non-singular on the half infinite line $(-\infty,\tau_{\rm max})$ 
		with some real  $\tau_{\rm max}$. In particular, this always happens when the couplings $r$ and $l$ are 
restricted as $-1<r(\tau),l(\tau)<0$.
		Such solutions of the Ricci flow equation,  
		which can be continued to infinite negative $\tau$, are called `ancient'. That \eqref{rff} admits  ancient solutions  suggests that the corresponding NLSM is a consistent  QFT. The  factorized scattering theory for the model was proposed  in ref.\cite{Fateev}.
		
		\bigskip
		
		The NLSM with  metric  \eqref{i389dw87} is an integrable classical field theory.  The  ZCR for the Euler-Lagrange equations
		was originally obtained in the work \cite{Lukyanov}. This way, the Fateev model provides an additional example of the link
		between  integrability and one-loop renormalizability in sigma models.

		\section{Poisson-Lie deformation \label{sec3}}
		The models discussed above  illustrate  the connection between  integrable
		NLSM and solutions of the Ricci flow equation. This can be used as a guiding principle for 
		constructing   new multiparametric families of metrics that satisfy the Ricci flow. Here we will discuss
		the so-called  Poisson-Lie deformation of integrable NLSM. 
		Such a deformation preserves integrability and
		allows one to obtain new solutions of the Ricci flow equation.
		We first  illustrate the idea by showing that the  anisotropic 
		${\rm SU}(2)$ PCF can be obtained as the Poisson-Lie deformation of the ${\rm SU}(2)$ PCF \cite{Klimchick3}. Then, 
		a new integrable model is constructed by deforming the fully anisotropic ${\rm SU}(2)$ PCF.

		\subsection{Poisson-Lie deformation of PCF}
		
		To explain the  Poisson-Lie deformation,  we start from the  Hamiltonian formulation of the model. The
		latter, in the case of the ${\rm SU}(2)$ PCF, can be described using the currents $J_i^a$ \eqref{re}.
		It follows from the Lagrangian \eqref{L2},\,\eqref{Ls} that they form a closed Poisson algebra \cite{Faddeev}:\footnote{%
			In our discussion of the Poisson-Lie
			deformation we set $e^2=1$. Since it appears in an  overall factor  multiplying
			the Lagrangian,   this has no effect on the classical equations of motion.}
		\bea \label{pb1}
		\{J_0^a(x),\, J_0^b(y) \}&=&\, \epsilon_{abc}\ J_0^c(x)\, \delta(x-y)\nonumber\\[0.2cm]
		\{J_0^a(x),\, J_1^b(y) \}&=& \epsilon_{abc}\ J_1^c(x)\, \delta(x-y)- \delta^{ab}\,\delta'(x-y)\\[0.2cm]
		\{J_1^a(x),\, J_1^b(y) \}&=&0\,.\nonumber
		\eea
		These are understood to be equal-time relations with $x^0=y^0$, while $x\equiv x^1$ and $y\equiv y^1$ are the space coordinates (the  dependence of the currents on the time variable has been suppressed). The Hamiltonian is obtained by means of the Legendre transform
		and is given by
		\be\label{asjk78we}
		H=\frac{1}{2}\int {\rd}x\,\sum\limits_{a=1}^3\big(J_{0}^a\,J_{0}^a+J_{1}^a\,J_{1}^a\big)\,.
		\ee
		One can check that the Hamiltonian equations of motion, $\dot O=\{H, O \}$, for the currents  are equivalent to 
eqs.\,\eqref{eqr} and \eqref{aaa} with $I_1=I_2=I_3$, i.e.,
		\be
		\p_- J_+^a=\tfrac{1}{2}\,\epsilon_{abc}\, J_+^bJ_-^c\,,\qquad\qquad \p_+ J_-^a=\tfrac{1}{2}\,\epsilon_{abc}\, J_-^bJ_+^c\,.
		\label{1}
		\ee
		\bigskip

		The Poisson algebra \eqref{pb1} admits a certain deformation which preserves its defining properties, namely,
		skew-symmetry, the Jacobi  and Leibniz identities. The deformed Poisson bracket relations read explicitly as
		\bea\label{3}
		\{\tilde J_{0}^a(x),\, \tilde J_{0}^b(y) \}&=&\, \frac{1}{1+r}\ \epsilon_{abc}\ \tilde J_{0}^c(x)\, \delta(x-y)\nonumber\\[0.2cm]
		\{\tilde J_{0}^a(x),\,\tilde J_{1}^b(y) \}&=&\frac{1}{1+r}\ \epsilon_{abc}\ \tilde J_{1}^c(x)\, \delta(x-y)- \delta^{ab}\,\delta'(x-y)\\[0.2cm]
		\{\tilde J_{1}^a(x),\,\tilde J_{1}^b(y) \}&=&\, -\frac{r}{1+r}\ \epsilon_{abc}\ \tilde J_{0}^c(x)\, \delta(x-y)\,.\nonumber
		\eea
		Here $r$ plays the role of the deformation parameter and we switch the notation from $J_i^a$ to $\tilde J_i^a$ as the above Poisson brackets will be associated with a different classical field theory. 
		Remarkably, with the same form of the Hamiltonian as \eqref{asjk78we}, i.e.,
		\be\label{h2}
		\tilde H=\frac{1}{2}\,\int {\rd}x\,\sum\limits_{a=1}^3\big(\tilde J_{0}^a\,\tilde J_{0}^a+\tilde J_{1}^a\,\tilde J_{1}^a\big)
		\ee
		the equations of motion do not depend on the deformation parameter. 
		Namely, they coincide with \eqref{1} 
		upon replacing $J_\pm^a$ by $\tilde J_\pm^a=\frac{1}{2}(\tilde J_0^a\pm \tilde J_1^a)$.
		This means that the Hamiltonian
		system defined through \eqref{3} and \eqref{h2} is integrable  by construction. The corresponding flat connection entering into the ZCR takes the same form as for the ${\rm SU}(2)$ PCF \eqref{zcr11} but written in terms of the currents $\tilde J^a_\pm$:
		\be\label{fc}
		{\bm A}_{\pm}=\frac{\ri\,\tilde J^a_{\pm}\,{\tt t}_a}{1\pm \lambda}\,.
		\ee
		The obtained  classical field theory  is called the Poisson-Lie deformation of the ${\rm SU}(2)$ PCF. 
		The final and  technically most involved step of the procedure is to derive the Lagrangian of the deformed model.
		\bigskip
		
		It is well known in classical mechanics how to get from the Hamiltonian to the Lagrangian picture. Consider a  mechanical system with a finite number of degrees of freedom $d$.
		The Poisson brackets are defined on the algebra of functions on the $2d$-dimensional phase space. In  local coordinates $(z^1,\ldots, z^{2d})$  they are given by
		\be
		\{f,g \}=\Omega^{AB}\,\frac{\p f}{\p z^A}\, \frac{\p g}{\p z^B}\, .
		\ee
		Since the Poisson brackets are assumed to be non-degenerate, the inverse of the contravariant tensor
		$\Omega^{AB}$ exists and we will denote it as $\Omega_{AB}$. Due to skew-symmetry of the Poisson brackets, the covariant tensor $\Omega_{AB}$ is antisymmetric, i.e.,
		it defines a two-form as $\Omega=\Omega_{AB}\,\rd z^A\wedge \rd z^B$.
		Moreover, the Jacobi identity implies that the form is closed, ${\rd}\Omega=0$. 
		This allows one to write  $\Omega$ as an exact form, 
		$\Omega={\rd}\alpha$, at least locally. The action is expressed in terms
		of  the one-form $\alpha$ and the Hamiltonian as
		\be\label{i128sd7}
		{ S}=\int \big(\alpha-H\, {\rd t} \big)
		\ee
		with the integral being taken over a path in the phase space parameterized by the time $t$. 
		According to the Darboux theorem there exists (locally) a set of canonical variables $(q^1,\ldots, q^d, p_1,\ldots p_d)$ such that 
		$\alpha=\sum_{m=1}^d p_m\,\rd q^m$. Then the Lagrangian 
		associated with the action \eqref{i128sd7} is given by
		\be\label{lag}
		L=\sum_{m=1}^d p_m\,\dot{q}^m-H\,.
		\ee
		This can be interpreted as the Legendre transform of $H$ where the canonical momenta $p_m$ are replaced by $\dot q^m$ as the independent variables. 
		\bigskip

		In order to apply the above procedure to the infinite dimensional Hamiltonian structure \eqref{3},\,\eqref{h2},
		it is useful to realize the Poisson algebra  in terms of the fields, similar to the canonical variables
		$p_m$ and $q^m$ in the finite dimensional case. For this reason we introduce local coordinates $X^\mu$ on the group manifold and  the corresponding 
		canonical momentum densities $\Pi_\mu$. They obey  the Poisson bracket relations
		\be\label{ak389}
		\{\Pi_\mu(x),X^\nu(y)\}=\delta_\mu^\nu\,\delta(x-y)\, ,\qquad\qquad
		\{X^\mu(x),X^\nu(y)\}=\{\Pi_\mu(x),\Pi_\nu(y)\}=0\,.
		\ee
		In the case $r=0$, when \eqref{3} becomes the undeformed algebra \eqref{pb1}, the currents 
\be
{\bm K}_i\equiv\tilde{J}_i^a\,{\tt t}_a\big|_{r=0}\qquad\qquad (i=0,1)
\ee
can be expressed in terms of the canonical fields in the following way; first, define the $3\times 3$ matrix $E^a\,_\mu$ through the relation
		\be
		{\rd}{\bm U}\,{\bm U}^{-1}= \ri\,E^a\,_{\mu}\,{\rd} X^\mu\,{\tt t}_a\,.
		\label{UU}
		\ee
		Its inverse will be denoted by $ E^{\mu a}$ so that $E^a\,_{\mu} E^{\mu b}=\delta^{a b}$. Then with the choice 
		\be\label{AA}
		{\bm K}_0= E^{\mu a}\,\Pi_\mu\,{\tt t}_a,\qquad\qquad\qquad  {\bm K}_1= E^a\,_\mu\,\p_1X^\mu\,{\tt t}_a=-\ri\,\p_1{\bm U}\,{\bm U}^{-1}
		\ee
		one can check via a direct computation that the Poisson algebra \eqref{pb1} with 
$J_i^a$ replaced by the components of ${\bm K}_i$ is satisfied. In fact, the r.h.s. of the first equation in \eqref{AA} 
is just $-\ri\,\p_0{\bm U}\,{\bm U}^{-1}$ written in terms of the canonical fields for the PCF.\footnote{We realise the algebra \eqref{pb1} using the `left' currents 
${\bm K}_i=-\ri\,\p_i{\bm U}\,{\bm U}^{-1}$ rather than the `right' ones $J^a_i{\tt t}_a=\ri\,{\bm U}^{-1}\,\p_i{\bm U}$  \eqref{re} for future convenience. 
The latter obey the same Poisson bracket relations \eqref{pb1}.}
		\bigskip
		
		For general $r\neq0$ one should first apply the
		linear transformation
		\be\label{310}
		{\cal I}^a=\frac{1+r}{2\sqrt{r}}\,\Big(\sqrt{r}\,\tilde J_0^a+\ri\,\tilde J_1^a\Big)\,,\qquad\qquad {\cal J}^a=\frac{1+r}{2\sqrt{r}}\,\Big(\sqrt{r}\,\tilde J_0^a-\ri\,\tilde J_1^a\Big)\,.
		\ee
		This brings the closed Poisson  algebra \eqref{3} to the form:
		\bea\label{currr}
		\{{\cal I}^a(x),\, {\cal I}^b(y) \}&=&\epsilon_{abc}\,{\cal I}^c(x)\, \delta(x-y)-k\, \delta_{ab}\,\delta'(x-y)\nonumber\\[0.2cm]
		\{{\cal J}^a(x),\, {\cal J}^b(y) \}&=&\epsilon_{abc}\,{\cal J}^c(x)\, \delta(x-y)+k\, \delta_{ab}\,\delta'(x-y)\\[0.2cm]
		\{{\cal I}^a(x),\, {\cal J}^b(y) \}&=& 0\,,\nonumber
		\eea
		where
		\be
		k=\ri\,\frac{(1+r)^2}{2\sqrt{r}}\,,
		\ee
		which is a direct sum of two independent so-called ${\rm SU}(2)$ current algebras. It turns out that the Poisson algebra generated by
		${\cal I}^a$ and ${\cal J}^a$ can be 
formally 
realised in terms of the currents $ {\bm K}_i$ \eqref{AA} as well as the group valued field ${\bm U}\in{\rm SU}(2)$.
		The explicit formula, along with its verification, is contained in ref. \cite{DMV} and is given by
		\bea\label{315}
		{\cal I}^a\, {\tt t}_a&=&\tfrac{1}{2}\,\big(1-\ri\, {\rm Ad}^{-1}_{\bm U}\circ R\circ {\rm Ad}_{{\bm U}} \big)\, 
		\bm{K}_0+k\,\bm{K}_1\nonumber\\[0.2cm] 
		{\cal J}^a\, {\tt t}_a&=&\tfrac{1}{2}\,\big(1+\ri\, {\rm Ad}^{-1}_{\bm U}\circ R\circ {\rm Ad}_{{\bm U}}\big)\, 
	\bm{K}_0-k\, \bm{K}_1\ .
		\eea
		Here
${\rm Ad}_{\bm U}$ stands for the adjoint action of the group, see   eq.\,\eqref{adj}, while the
  linear operator $R: \mathfrak{su}(2)\mapsto\mathfrak{su}(2)$  
is defined via its action on the generators as
		\be\label{RRR}
		R({\tt t}_1)={\tt t}_2\,,\qquad R({\tt t}_2)=-{\tt t}_1\,,\qquad  R({\tt t}_3)=0\,.
		\ee
		\bigskip
		
		Formulae \eqref{310}, \eqref{315} and \eqref{AA} allow one to realize the currents $\tilde{J}_0^a$ and $\tilde{J}_1^a$, satisfying the Poisson bracket relations
		\eqref{3},  through the canonical fields  \eqref{ak389}. The corresponding expression for the Hamiltonian follows from \eqref{h2}.
		In the basis of canonical variables the construction of the Lagrangian is straightforward and is the
		field theory analogue of the Lengedre transform \eqref{lag}. Applying the procedure, where 
		$\Pi_\mu$ maps to $\dot X^\mu=\{\tilde{H}, X^\mu\}$, one arrives at the  Lagrangian density
		\be\label{ooo}
		{\cal L}=-\frac{1+r}{e^2}\,\big\langle {\bm U}^{-1}\,\p_+{\bm U}\,, {\cal O}\,\big( {\bm U}^{-1}\, \p_{-}{\bm U}\big)\big\rangle\,\qquad\qquad {\rm with}\qquad\qquad{\cal O}=\big(1-\sqrt{r}\, R \big)^{-1}\,.
		\ee
		Here the dependence on $e^2$ was restored and we performed the substitution $e^2\mapsto (1+r)\,e^2$ to keep with the conventions of  section \ref{sec1}.
		\bigskip
		
		At first glance, in local coordinates, $\cal L$  can not be written in the form \eqref{Ls}. Instead, the latter should be  modified as
		\be\label{uyt}
		{\cal L}=2\,G_{\mu\nu}(X)\,\p_+X^\mu\,\p_-X^\nu-B_{\mu\nu}(X)\, \big(\p_+X^\mu\,\p_-X^\nu-\p_-X^\mu\,\p_+X^\nu \big)\,.
		\ee
		Here the last term  is not invariant w.r.t. the   parity transformation $x^1\mapsto -x^1$, i.e., $\partial_\pm\mapsto \partial_\mp$ and comes about because the Lagrangian density \eqref{ooo} is not either.
		Models of this type motivate a 
		generalization of the NLSM where the target space is additionally equipped 
		with a two form  $B=B_{\mu\nu}\,{\rd}X^\mu\wedge{\rd}X^\nu$  known as the $B$-field \cite{Braaten}. It turns out
		that in the ${\rm SU}(2)$ case the $B$-field corresponding to ${\cal L}$ \eqref{ooo} is a closed form
(in fact, exact).
		As a result, the term $\propto B_{\mu\nu}$ in \eqref{uyt} is a total derivative and 
		has no effect on the Euler-Lagrange equations.  This way,  for the ${\rm SU}(2)$ case,
		the obtained sigma model is equivalently described by  \eqref{Ls} where
		\be\label{saj9821321}
		G_{\mu\nu}(X)\,{\rd}X^\mu{\rd}X^\nu=-\frac{1+r}{e^2}\,\big\langle {\bm U}^{-1}\,{\rd}{\bm U}\,, {\cal O}_{{\rm sym}}\,( {\bm U}^{-1}\,{\rd}{\bm U})\big\rangle \qquad {\rm with}\qquad {\cal O}_{{\rm sym}}=\big(1+r-r\, P_3  \big)^{-1}
		\ee
and $P_3=1+R^2\in{\rm End}\big(\mathfrak{su}(2) \big)$ stands for  the projector on the Cartan subalgebra generated by ${\tt t}_3$. 
		This way we arrive at the metric of the anisotropic ${\rm SU}(2)$ PCF \eqref{apfc}.
		
		\bigskip
		
		It was discussed in section \ref{sec1} that the anisotropic ${\rm SU}(2)$ PCF is a  integrable classical field theory. Having established that the model is a Poisson-Lie deformation of the ${\rm SU}(2)$ PCF, we obtain a way to derive the Zero Curvature Representation
 for the classical equations of motion. Namely, the flat connection   is given by \eqref{fc} where the currents $\tilde J_{\pm}^a=\frac{1}{2}(\tilde J_{0}^a\pm\tilde J_{1}^a)$ entering therein
read as
		\be\label{tttj}
		\tilde J_{\pm}^a{\tt t}_a=(1+r)\,{\rm Ad}_{\bm U}\circ\,\big(1\pm \sqrt{r}\,R\big)\,\big(\p_{\pm}{\bm U}\, {\bm U}^{-1}\big)\,.
		\ee
		Indeed, as it follows from the Euler-Lagrange equations for the model \eqref{ooo},
		\be
		\p_- \tilde J_+^a=\tfrac{1}{2}\,\epsilon_{abc}\, \tilde J_+^b\tilde J_-^c\,,\qquad\qquad \p_+ \tilde J_-^a=\tfrac{1}{2}\,\epsilon_{abc}\, \tilde J_-^b\tilde J_+^c\,.
		\ee
		\bigskip
		
		The following comment is in order here.  The anisotropic ${\rm SU}(2)$  PCF admits an integrable generalization, where ${\bm U}$ belongs to an arbitrary  simple Lie group $G$.
		The Lagrangian is still given by \eqref{ooo} with $R$
    		being a certain linear operator which is usually referred to as the
Yang-Baxter operator. It acts on the Lie algebra $\mathfrak{g}={\rm Lie}(G)$ and
		is required to satisfy  a skew symmetry condition and the so-called
		modified Yang-Baxter equation \cite{STS}. A possible choice obeying the two properties is specified
		using the Cartan-Weyl decomposition of the simple Lie algebra, 
		$\mathfrak{g}=\mathfrak{n}_+\oplus \mathfrak{h}\oplus\mathfrak{n}_-$, where $\mathfrak{h}$ stands for the Cartan subalgebra and $\mathfrak{n}_\pm$ are the nilpotent ones. Namely, 
the linear operator $R$ is unambiguously defined through the conditions
		\be\label{sakj983}
		R ({\tt e}_{\pm})=\mp\ri \,{\tt e}_{\pm}\,,\qquad\qquad\qquad R({\tt h})=0\qquad\qquad \big(\forall {\tt e}_{\pm}\in\mathfrak{n}_\pm,\,\forall\,{\tt h}\in\mathfrak{h}\big)\,.
		\ee 		The NLSM  \eqref{ooo} with $R$ being the Yang-Baxter operator
was introduced by Klim\c{c}\'{i}k in ref.\,\cite{Klimcik6} who called it the
Yang-Baxter sigma model. Written in terms of local coordinates, the Lagrangian
		takes the form \eqref{uyt} where, for  general group,
		the second term $\propto B_{\mu\nu}$   is no longer a total derivative and cannot be ignored. 
		The model is classically integrable and the corresponding 
flat connection is given by the same formulae \eqref{fc} and \eqref{tttj} \cite{Klimchick3}.
		
		\bigskip
		
		The Yang-Baxter sigma model also turns out to be a one-loop renormalizable field theory. The proof is based on the  
		extension of the results of the works \cite{EH, Friedan} to the case of an  NLSM equipped with a $B$-field that was carried out in ref.\cite{Braaten}, see also the textbook \cite{GSW}.
		The one-loop RG flow equations are modified from \eqref{ricci} as
		\bea\label{wersfecd}
		\p_\tau G_{\mu\nu}&=&-\hbar\,\big(R_{\mu\nu}-\tfrac{1}{4}H_{\mu}\ ^{\sigma \rho}H_{\sigma \rho \nu}\big)+O(\hbar^2)\nonumber\\[0.2cm]
		\p_\tau B_{\mu\nu}&=&-\tfrac{1}{2}\,\hbar\,\nabla_\sigma H^\sigma\ _{\mu\nu}+O(\hbar^2)\,.
		\eea
		Here $H_{\mu\nu\lambda}$ are the components of the so-called torsion tensor. It is given by the exterior derivative of the $B$-field, i.e.,
		\be\label{askj893jh21as}
		H_{\mu\nu\lambda}=\p_\mu\, B_{\nu\lambda}+\p_\nu\, B_{\lambda\mu}+\p_\lambda\, B_{\mu\nu}\,.
		\ee
		For the model \eqref{ooo}, \eqref{uyt} with $\bm{U}$ belonging to a simple Lie group,
 the above equations boil down to a system of ordinary differential equations on $e^2$ and $r$. They read as 
\cite{Squellari}
		\bea\label{sakj8jassasa}
  -\frac{1}{\hbar}\,\p_\tau(e^{-2})&=&\tfrac{1}{2}\,C_2\,(1-r)\,\nonumber\\[0.2cm]
		-\frac{1}{\hbar}\,\p_\tau r &=&C_2\,e^2 r \,(r+1)\,,
		\eea
		where, remarkably, the only dependence on the group appears through an overall factor proportional to the value of the quadratic Casimir in the adjoint representation. Note that in the domain $-1<r<0$,
for which the system \eqref{sakj8jassasa} possesses ancient solutions, the deformation parameter $\sqrt{r}$
entering into the Lagrangian of the Yang-Baxter sigma model \eqref{ooo} is an imaginary number. 
The corresponding target-space 
metric \eqref{saj9821321} remains real. However, the torsion tensor \eqref{askj893jh21as},
which is non-vanishing outside the ${\rm SU}(2)$ case,
becomes purely imaginary 
(a related discussion is contained in Appendix A of ref.\cite{Gleb}).

		\bigskip

\begin{figure}[]
			\begin{center}
				\begin{tikzpicture}
					\node[draw,fill=blue!13,rounded corners=3,align=center] at (-2, 0) {${\rm SU}(2)$ PCF};
					\draw [line width=0.5mm,->] (-0.5,0)--(2,0);
					\node[draw, text width=2cm, fill=blue!13,rounded corners=3,align=center] at (3.5,0) {anisotropic ${\rm SU}(2)$ PCF};
					\draw [line width=0.5mm,->] (5,0)-- (7.5,0);
					\node (FM) [draw,text width=2.5cm,fill=blue!13,rounded corners=3,align=center] at (9.3,0) {Fateev model};
				\end{tikzpicture}
				\caption{\small%
The relation between the various models. The Poisson-Lie deformation is represented by an arrow.  }

			\label{f7}
		\end{center}
	\end{figure}
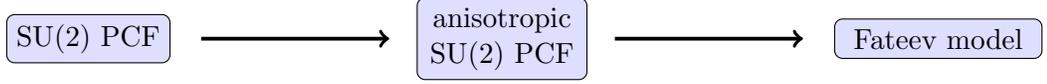

		We have just discussed that the  Poisson-Lie deformation  of the  PCF yields the Yang-Baxter sigma model. The latter itself can be deformed along the 
similar line of arguments \cite{Klimchick3}, see also \cite{Gleb} as well as fig.\,\ref{f7} for a summary. 
In the case of $G={\rm SU}(2)$ the
		obtained theory turns out to be the Fateev model, i.e., the sigma model with target space metric \eqref{i389dw87}. 
		For a general simple Lie group $G$, the two parameter deformation of the PCF
was introduced by Klim\v{c}\'{i}k  in ref.\cite{Klimchick3}. The corresponding Lagrangian
		involves the Yang-Baxter operator $R$ and is given by
	\be\label{askj98123}
		{\cal L}=-\frac{2(1+r)(1+l)}{e^2}\
		\big\langle {\bm U}^{-1}\,\p_+{\bm U}\,, {\cal O}\,\big( {\bm U}^{-1}\, \p_{-}{\bm U}\big)\big\rangle\,
		\ee
		with 
		\be
		{\cal O}=\Big(1-\sqrt{l(1+r)}\ R-\sqrt{r(1+l)}\ {\rm Ad}_{\bm U}\circ R\circ{\rm Ad}_{\bm U}^{-1}  \Big)^{-1}\,.
		\ee
		For $\bm{U}\in{\rm SU}(2)$
		the $B$-field turns out to define a closed two form and has no effect on the equations of motion. It was shown in \cite{Hoare} by an explicit computation that the
		metric is equivalent to \eqref{i389dw87}.
		For arbitrary simple Lie group $G$ the model  \eqref{askj98123}  is classically integrable and the ZCR was found in ref.\,\cite{Klimchick4}. One-loop renormalizability
		was demonstrated in the work \cite{SST} using the results of \cite{VKQ}. The differential equations describing the
		flow of the couplings $(e^2,r,l)$ are
  \bea
  -\frac{1}{\hbar}\,\p_\tau(e^{-2})&=&\frac{ C_2\, (1-l-r) (l+r+1)}{2\,(1+l) (1+r)}\,\nonumber\\[0.2cm]
  -\frac{1}{\hbar}\,\p_\tau r&=&\frac{C_2\,e^2\, r \,  (l+r+1)}{1+l}\\[0.2cm]
-\frac{1}{\hbar}\,\p_\tau l&=&\frac{C_2\,e^2\, l \,  (l+r+1)}{1+r}\,.\nonumber
  \eea
They essentially  coincide with  \eqref{rff} which were derived in Fateev's original paper \cite{Fateev}.

	\subsection{Poisson-Lie deformation of fully anisotropic ${\rm SU}(2)$ PCF\label{sec32}}
Here we obtain a
 new clasically integrable NLSM as a Poisson-Lie deformation of the fully anisotropic ${\rm SU}(2)$ PCF.
The procedure closely follows that which was explained above  on the example of the ${\rm SU}(2)$ PCF. 
\bigskip

The Hamiltonian for the fully anisotropic ${\rm SU}(2)$ PCF \eqref{Ls}, \eqref{mm1}, written in  terms of the currents \eqref{re},
is given by
 \be\label{hhh}
	H=\frac{1}{2}\,\sum_{a=1}^3\int {\rd x}\, I_a\,\Big(J_0^aJ_0^a+ J_1^aJ_1^a \Big)\,,
	\ee
while the equal-time Poisson bracket relations for $J_i^a$ read as
 \bea \label{pb6}
	\{J_0^a(x),\, J_0^b(y) \}&=&\,\frac{I_c}{I_a\, I_b}\, \epsilon_{abc}\ J_0^c(x)\, \delta(x-y)\nonumber\\[0.2cm]
	\{J_0^a(x),\, J_1^b(y) \}&=& \frac{1}{I_a}\,\epsilon_{abc}\ J_1^c(x)\, \delta(x-y)-\frac{1}{I_a}\, \delta^{ab}\,\delta'(x-y)\\[0.2cm]
	\{J_1^a(x),\, J_1^b(y) \}&=&0\,.\nonumber
	\eea
The above Poisson algebra admits  a deformation of the form
\bea \label{tildeJs}
\{\tilde J_0^a(x),\, \tilde J_0^b(y) \}&=&\,\frac{I_c-\xi}{I_a\, I_b}\ \epsilon_{abc}\  \tilde J_0^c(x)\, \delta(x-y)\nonumber\\[0.2cm]
\{\tilde J_0^a(x),\, \tilde J_1^b(y) \}&=&\frac{I_b-\xi}{I_a\,I_b}\ \epsilon_{abc}\  \tilde J_1^c(x)\, \delta(x-y)-\frac{1}{I_a}\, \delta^{ab}\,\delta'(x-y) \\[0.2cm]
\{\tilde J_1^a(x),\,\tilde J_1^b(y) \}&=&\, -\frac{\xi}{I_a\,I_b}\ \epsilon_{abc}\ \tilde J_{0}^c(x)\, \delta(x-y)\nonumber
\eea
depending on the extra parameter $\xi$.
Then, with the Hamiltonian
\be\label{askj23874vhbdnj}
	\tilde H=\frac{1}{2}\,\sum_{a=1}^3\int {\rd x}\, I_a\,\Big(\tilde{J}_0^a\tilde{J}_0^a+ \tilde{J}_1^a\tilde{J}_1^a \Big)\,,
	\ee
which is formally the same as \eqref{hhh} but expressed in terms of the new  currents $\tilde J_i^a$, the Hamiltonian equations of motion imply
\bea\label{kjas893}
	\p_-\tilde J_+^a&=&\frac{I_a+I_b-I_c}{2I_a}\ \tilde J_+^b\,\tilde J_-^c-\frac{I_a-I_b+I_c}{2I_a}\ \tilde J_+^c\,\tilde J_-^b\nonumber\\[0.2cm]
	\p_+\tilde J_-^a&=&\frac{I_a+I_b-I_c}{2I_a}\ \tilde J_-^b\,\tilde J_+^c-\frac{I_a-I_b+I_c}{2I_a}\ \tilde J_-^c\,\tilde J_+^b\, .
	\label{jh}
	\eea
Here $(a,b,c)={\tt perm}(1,2,3)$ and summation over repeated indices is not being assumed.
The equations \eqref{kjas893} are equivalent to  \eqref{eqr},\,\eqref{aaa} up to the replacement $J_i^a\mapsto \tilde{J}_i^a$.
\bigskip

The currents $\tilde{J}_i^a$ obeying the Poisson bracket relations \eqref{tildeJs} can be realized in terms of the  fields
$X^\mu$ and $\Pi_\mu$ subject to the canonical commutation relations \eqref{ak389}.  
This is done along the same line of arguments as was discussed in the previous subsection. Namely,  one first considers
certain linear combinations of $\tilde{J}_i^a$ which 
obey two independent copies of the classical ${\rm SU}(2)$ current algebra \eqref{currr} with $k$ being a certain function of the couplings
$I_a$ and deformation parameter $\xi$.
Then realizing ${\cal I}$ and ${\cal J}$ in terms of the
canonical variables (see formulae \eqref{315},\,\eqref{AA}) and
performing the Legendre transform
of the Hamiltonian \eqref{askj23874vhbdnj}, one obtains the Lagrangian of the deformed theory.
The result of the calculations reads as
\be\label{kjsa9812j}
{\cal L}=-4\left\langle\, {\bm U}^{-1}\partial_-{\bm U}\,,\, {\cal O}_+\,\big( {\bm U}^{-1}\partial_+{\bm U}\big) \right\rangle\, ,
\ee
where a certain choice of the overall multiplicative factor for the Lagrangian density was made.
Here  and below we use the notation ${\cal O}_{\pm}$ for the linear operators acting on the Lie algebra $\mathfrak{su}(2)$  given by
\be\label{asjk8973BBBB}
{\cal O}_{\pm}=\bigg(\frac{1}{I_1-\xi}\, P_1+\frac{1}{I_2-\xi}\, P_2+\frac{1}{I_3-\xi}\, P_3\pm \sqrt{\gamma}\,{\rm Ad}_{\bm U}\circ R\circ{\rm Ad}^{-1}_{\bm U} \bigg)^{-1}
\ee
with
\be\label{gamma1}
\gamma= \frac{\xi}{(I_1-\xi)(I_2-\xi)(I_3-\xi)}\,.
\ee
\bigskip

The Lagrangian density \eqref{kjsa9812j} is formally not invariant under the parity transformation $x^1\mapsto - x^1$ (so that $\p_\pm\mapsto\p_\mp$). 
Nevertheless, the theory possesses this symmetry. The reason is because in local coordinates, where  ${\cal L}$ \eqref{kjsa9812j} 
takes the form \eqref{uyt}, the term $\propto B_{\mu\nu}$ turns out to be a total derivative. Thus one is free to replace ${\cal O}_+$ in \eqref{kjsa9812j} by 
\be
{\cal O}_{\rm sym}=\frac{1}{2}\,\big({\cal O}_++{\cal O}^{\rm T}_+\big)\,, 
\ee
where the transposition is defined by the condition 
$\langle {\tt x}\,, {\cal O}_+\,{\tt y}\rangle=\langle {\cal O}^{{\rm T}}_+ \,{\tt x}\,, {\tt y}\rangle$  for any
 ${\tt x}, {\tt y}\in {\mathfrak{su}(2)}$.
This way, the target space metric for the deformed sigma model can be written as
\be\label{asjk8973AAAA}
G_{\mu\nu}(X)\,{\rd}X^\mu{\rd}X^\nu=- 2\,\big\langle {\bm U}^{-1}\,{\rd}{\bm U}\,,{\cal O}_{\rm sym}( {\bm U}^{-1}\,{\rd}{\bm U})\big\rangle\,.
\ee
It is worth mentioning that for $I_1=I_2$ this becomes the Fateev  metric \eqref{i389dw87},\,\eqref{i389dw87AA} upon the identification of parameters:
\be
I_1=I_2=\frac{(1+l)^2}{2e^2}\,,\qquad\qquad I_3=\frac{(1+l)\,(1+l+r)}{2e^2}\,,\qquad\qquad \xi=\frac{l\,(l+1)}{2e^2} \ .
\ee
\bigskip

By construction 
the obtained model \eqref{kjsa9812j} is a classically integrable field theory. The
corresponding flat connection takes the same form as for the fully anisotropic ${\rm SU}(2)$ PCF, i.e.,
\be\label{aaaaasj783hjsd}
	{\bm A}_\pm=\ri\,\sum\limits_{a=1}^3w_a(\nu\mp\lambda)\tilde{J}_\pm ^a\, {\tt t}_a\,,
	\ee
 where the functions $w_a(\lambda)$ are given in \eqref{mpm} and \eqref{wwwww}. The formula for the currents $\tilde J^a_\pm$ in terms of the ${\rm SU}(2)$ element ${\bm U}$ reads as
\be
	\tilde J^a_\pm= \ri\,C_a\,\big\langle {\cal O}_{\pm}\,\big( {\bm U}^{-1}\partial_{\pm}{\bm U}\big), {\tt t}_a\big\rangle\,, \qquad\qquad
 C_a=\frac{2}{I_a-\xi}\ \sqrt{\frac{I_b\, I_c}{(I_b-\xi)\,(I_c-\xi)}}\,
	\ee
 with $(a,b,c)={\tt perm}(1,2,3)$.
 \bigskip

One can check that
the metric \eqref{asjk8973BBBB}-\eqref{asjk8973AAAA}
satisfies the Ricci flow equation \eqref{ricci}. The parameter $\gamma$ defined in \eqref{gamma1} turns out to be an RG invariant, i.e.,
\be\label{gamma33}
\frac{1}{\hbar}\,\p_\tau\gamma=O(\hbar)\,.
\ee
As for the couplings $I_a=I_a(\tau)$, it is convenient to swap these in favour of  $\tilde I_a$ according to
 \be
 \tilde I_a\equiv I_a-\xi\,.
 \ee
The latter obey the RG flow equations
	\be\label{sys22}
	-\frac{1}{\hbar}\,\p_\tau\big(\tilde I_a\,\tilde I_b\big)=
\big(1+\gamma\, \tilde I_a\, \tilde I_b \big)\, \big(\tilde I_a+\tilde I_b-\tilde I_c+\gamma\,\tilde I_a\,\tilde I_b\,\tilde I_c \big)+O(\hbar)\,,\qquad\qquad (a,b,c)={\tt perm}(1,2,3)\,,
	\ee
which may be compared to the underformed case \eqref{sakj39821}.
The derivation of the above formulae uses the property that the Ricci tensor corresponding to the metric \eqref{asjk8973BBBB}-\eqref{asjk8973AAAA} can be written as
	\be
R_{\mu\nu}=\sum\limits_{a=1}^3\frac{(\tilde I_a-\tilde I_b+\tilde I_c+\gamma\, \tilde I_a\, \tilde I_b\, \tilde I_c ) (\tilde I_a+\tilde I_b-\tilde I_c+\gamma\, \tilde I_a\, \tilde I_b\, \tilde I_c )}{2\tilde I_b\, \tilde I_c }\,
\bigg(\frac{\p G_{\mu\nu}}{\p \tilde I_a}\bigg)_{\gamma}\  ,
	\ee
which generalizes the relation \eqref{akjs87hjd}.
 \bigskip
 
For $\gamma=0$ the two first integrals of \eqref{sys22} coincide with $Q_1$, $Q_2$  
\eqref{an8973jsd}\,-\,\eqref{an8973jsdC} with $I_a\mapsto \tilde{I}_a$. We found that these RG invariants
admit a deformation to arbitrary $\gamma$.  The explicit expressions involve, apart from
\be
\tilde{m}=\frac{\tilde{I}_2\,(\tilde{I}_1-\tilde{I}_3)}{\tilde{I}_3\,(\tilde{I}_1-\tilde{I}_2)}\,,\qquad\qquad\qquad
\tilde{p}=\frac{\tilde{I}_1}{\tilde{I}_2}\,,
\ee
also $m$ \eqref{an8973jsdC}, which enters into the functions
$w_a$ that appear in the flat connection \eqref{aaaaasj783hjsd}.
In terms of the parameters $\tilde{I}_a$, it is given by
\be
 m=
\frac{\tilde{I}_2\, (\tilde{I}_1-\tilde{I}_3)\,(1+\gamma\, \tilde{I}_1\,\tilde{I}_3) }{%
\tilde{I}_3\, (\tilde{I}_1-\tilde{I}_2)\,(1+\gamma\,\tilde{I}_1\, \tilde{I}_2)}\,.
\ee
The two first integrals of the system \eqref{sys22} read as
\bea\label{kasjyuhg218732}
Q_{1}^{(\gamma)}&=&\frac{K(1- m)-(1-\tilde{p})\,\tilde{m}\, \Pi(1-\tilde{m},1-m)}{(1-\tilde{p})(1-\tilde{m})\, 
\Pi(\tilde{m},m)+\tilde{p}\,K( m)} \nonumber\\[0.2cm]
Q_{2}^{(\gamma)}&=&\frac{ \tilde{m}-m}{\,\gamma\,}\ 
\frac{\big((\tilde{p}-1)(1-\tilde{m})\,\Pi(\tilde{m},m) -\tilde{p}\,K(m)\big)^2}{(\tilde{p}-1)^2\,\tilde{m}\,(1-\tilde{m})}\,,
\eea
where $\Pi(\tilde m, m)$ is the complete elliptic integral of the third kind:
\be
\Pi(\tilde m, m)=\int\limits_0^1\frac{\rd t}{(1-\tilde{m}\,t^2)\sqrt{(1-t^2)(1- m\,t^2)}}\,.
\ee
It is straightforward to check that $Q^{(0)}_1=Q_1$, while
$
\lim_{\gamma\to 0} Q^{(\gamma)}_2=Q_2
$.

\section{Summary and discussion}	
In this work we explored the interplay between integrability and one-loop renormalizability for NLSM in  $1+1$ dimensional spacetime. 
Our main example was the fully anisotropic ${\rm SU}(2)$ PCF.
On the one hand, it was explained that this is a clasically integrable field theory and the Zero Curvature Representation for
 its equations of motion was reviewed. On the other, the corresponding target space metric satisfies the Ricci flow  equation
\eqref{ricci} so that the  fully anisotropic ${\rm SU}(2)$ PCF  is one-loop
renormalizable within a three dimensional space of couplings. The system of ODEs describing the flow was derived and 
its full set of  first integrals was obtained,  independently from \cite{Halphen1,Halphen2}.
\bigskip

Another main 
result is the construction of a 
classically integrable  NLSM depending on four parameters
  whose Lagrangian density is given by \eqref{kjsa9812j}-\eqref{gamma1}. 
It was found by applying a Poisson-Lie deformation to the fully anisotropic ${\rm SU}(2)$ PCF. 
The corresponding 
target space metric turned out to provide a new solution to   the Ricci flow equation. 
The first integrals to the system of ODEs \eqref{gamma33} and \eqref{sys22}, which describe the flow of the 
four couplings, were derived in the course of this work and are given in \eqref{kasjyuhg218732}.

\bigskip

The class of theories that we discussed admit a modification such 
that they   remain one-loop renormalizable. This is achieved by adding the so-called Wess-Zumino term to the action. 
The Lagrangian takes the form \eqref{uyt} with the $B$-field no longer being  exact. This implies that  the target space, together with the Riemannian metric $G_{\mu\nu}$, is equipped with the affine
connection, where the torsion $H=\rd B$ is non-vanishing \cite{Braaten}. 
In the case of ${\rm SU}(2)$, the 3-form $H$ is proportional to the volume form for the group and can be written as 
\be\label{bbb}
H\equiv \rd B=\frac{{\tt k}}{24\pi}\,\Big\langle \big[{\bm U}^{-1}\,{\rd {\bm U}}\ \overset{\wedge}{,}\ {\bm U}^{-1}\,{\rd {\bm U}}\big]\ \overset{\wedge}{,}\ {\bm U}^{-1}\,{\rd {\bm U}}\Big\rangle\,.
\ee
Here ${\tt k}$ is an additional parameter of the model.  In the classical theory there is no contraint on
the values it may take, however, upon quantization it is required to be an RG invariant and, furthermore, must be an integer \cite{Witten1}.
For the case of the fully anisotropic ${\rm SU}(2)$ PCF with Wess-Zumino term,  the  
one-loop RG flow equations \eqref{wersfecd}
imply the system of ODEs for the couplings:
\bea  \label{sysWZ}
-\frac{1}{\hbar}\,\p_{\tau}(I_a\,I_b)&=&I_a+I_b-I_c-\frac{{\tt k}^2}{64\pi^2 I_c}+O(\hbar)\,,\qquad\qquad (a,b,c)={\tt perm}(1,2,3)\nonumber\\[0.2cm] 
 \frac{1}{\hbar}\,\p_\tau {\tt k}&=&0\,.
\eea
It possesses two Liouvillian first integrals, which are a simple generalization of \eqref{an8973jsd} and
in terms of $p$ and $m$  \eqref{an8973jsdC} take the form
	\bea
	Q_1&=&\frac{K(1-m)-(1-p) E(1-m)}{(1-p) E(m)+p K(m)}\,, \\[0.2cm]
  Q_2&=&\frac{I_1^2\,\big((p-1)E(m)-pK(m)\big)^2}{p\,(p-1)(pm-m+1)}\,+
\frac{{\tt k}^2}{64\pi^2}\,\frac{K(m)}{p-1}\,\big((p-1)E(m)-pK(m)\big)\ .
	\eea
 A complete analysis of the behaviour of the solutions to \eqref{sysWZ}
has not been carried out yet. Moreover, the classical integrability of the model has not been  established and the Zero Curvature Representation,
if it exists, remains unknown to us. These would
be interesting questions to pursue in future work. They can also be  addressed for the Poisson-Lie deformed theory.

\bigskip
Our work was mainly focused on sigma models associated with the Lie group ${\rm SU}(2)$. Nevertheless,
we expect it to be possible to
generalize the Poisson-Lie deformed theory constructed here to the case of higher rank Lie groups. 
One way to approach the problem uses the results of  ref.\cite{Lacroix:2023qlz}. In that paper,
a classically integrable NLSM is introduced, which is  a two parameter deformation of the PCF
for Lie group ${\rm SL}(N)$. 
For $N=2$ it coincides with the
fully anisotropic ${\rm SU}(2)$ PCF 
(upon an appropriate choice of reality conditions on the fields and parameters). 
We expect that  this sigma model may also be deformed along the  line of  arguments presented
in sec.\,\ref{sec3}. Another possibility for constructing integrable deformations,
based on the formalism of the so-called affine Gaudin model, 
 is mentioned in the perspectives section of  ref.\cite{Lacroix:2023qlz}.
\bigskip

Finally, classically integrable multiparametric families of sigma models 
are of interest to string theory. In particular, 
the possibility of an integrable elliptic deformation of strings on ${\rm Ad}_3\times {\rm S}^3\times {\rm T}^4$
was investigated in the recent paper \cite{Hoare:2023zti}.

\section*{Acknowledgments}
The authors  would like to thank Sergei Lukyanov for collaboration at the early stages of
this work and for his continued interest and support. GK   acknowledges discussions with  Sylvain Lacroix.
\smallskip

\noindent
Part of this work was carried out during GK's visits to the NHETC at Rutgers University. 
He is grateful for the support and hospitality he received during the stays.
\smallskip

\noindent
The research of DS is supported by the NSF under grant number NSF-PHY-2210187.

\end{document}